\documentclass[amsmath,amssymb,aps,prl,reprint,groupedaddress]{revtex4-1}

\usepackage{amsmath}
\usepackage{xcolor}
\usepackage{amsfonts,amssymb}
\usepackage{bm}
\usepackage{graphicx}% Include figure files
\begin{document}

%\preprint{}

\title{Grand-canonical Monte-Carlo simulation methods for charge-decorated cluster expansions}


\author{Fengyu Xie}
\email[]{fengyu\_xie@berkeley.edu}
%\thanks{}
\affiliation{Department of Materials Science and Engineering, University of California, Berkeley, California 94720, United States}
\affiliation{Materials Sciences Division, Lawrence Berkeley National Laboratory, California 94720, United States}

\author{Peichen Zhong}
\affiliation{Department of Materials Science and Engineering, University of California, Berkeley, California 94720, United States}
\affiliation{Materials Sciences Division, Lawrence Berkeley National Laboratory, California 94720, United States}

\author{Luis Barroso-Luque}
\affiliation{Department of Materials Science and Engineering, University of California, Berkeley, California 94720, United States}
\affiliation{Materials Sciences Division, Lawrence Berkeley National Laboratory, California 94720, United States}

\author{Bin Ouyang}
\affiliation{Department of Materials Science and Engineering, University of California, Berkeley, California 94720, United States}
\affiliation{Materials Sciences Division, Lawrence Berkeley National Laboratory, California 94720, United States}

\author{Gerbrand Ceder}
\email[]{gceder@berkeley.edu}
%\thanks{}
\affiliation{Department of Materials Science and Engineering, University of California, Berkeley, California 94720, United States}
\affiliation{Materials Sciences Division, Lawrence Berkeley National Laboratory, California 94720, United States}

%\date{\textcolor{red}{\today, Version 0}}

\begin{abstract}
Monte-Carlo sampling of lattice model Hamiltonians is a well-established technique in statistical mechanics for studying the configurational entropy of crystalline materials. When species to be distributed on the lattice model carry charge, the charge balance constraint on the overall system prohibits single-site Metropolis exchanges in MC. In this article, we propose two methods to perform MC sampling in the grand-canonical ensemble in the presence of a charge-balance constraint. The table-exchange method (TE) constructs small charge-conserving excitations, and the square-charge bias method (SCB) allows the system to temporarily drift away from charge neutrality. We illustrate the effect of internal hyper-parameters on the efficiency of these algorithms and suggest practical strategies on how to apply these algorithms to real applications.
\end{abstract}

%\keywords{}

\maketitle
\section{Introduction}
Configurational disorder is particularly important for understanding the thermodynamic properties of materials at non-zero temperatures, especially in systems composed of multiple components. The cluster-expansion (CE) method has been a successful approach to study the statistical mechanics of configurational disorder in solids\cite{de1979configurational,sanchez1984generalized, de1994cluster,ceder1993derivation}, and has been used to calculate phase diagrams in alloys\cite{kohan1998computation, van2002automating, ghosh2008first, ravi2012first} and ionic solids \cite{tepesch1996model, zhou2006configurational, avdw2010cluster, richards2018fluorination}, predict the short-range order related properties under finite temperatures\cite{wolverton1994long, wolverton2000short, seko2006first, ji2019hidden}, find the ground-state ordering in alloys\cite{zunger2001first, seko2006prediction, barabash2006prediction, seko2014efficient, huang2016finding, larsen2017alloy}, and even compute voltage profile of battery electrode materials\cite{aydinol1997ab, wolverton1998first, van1998first, y2002first, malik2009phase}. 

The CE model can be understood as a generalization of the Ising model. The micro-states in a solid solution are represented as a series of occupancy variables $\bm{\sigma}$, which denote the chemical species occupying each lattice site. The energy of a micro-state is described as a function of occupancy and is expanded as a sum of many-body interactions:
\begin{equation}
	E(\bm{\bm{\sigma}}) = \sum_{\bm{\beta}} m_{\bm{\beta}} J_{\bm{\beta}}\left\langle \Phi_{\bm{\alpha}}(\bm{\sigma}) \right\rangle_{\bm{\alpha} \in \bm{\beta}},
	\label{eq:CE}
 \end{equation}
where $\Phi_{\bm{\alpha}}$'s are a set of cluster basis functions that take as input the occupancy values of different clusters of multiple sites. The cluster basis functions are then grouped and averaged over lattice symmetry orbits $\bm{\beta}$ to generate the correlation functions $\left\langle \Phi_{\bm{\alpha}} \right\rangle_{\bm{\alpha} \in \bm{\beta}}$; and $m_{\bm{\beta}}$ is the multiplicity of orbit $\bm{\beta}$ per crystallographic unit cell.  The linear-expansion coefficients $J_{\bm{\beta}}$ are called effective cluster interactions (ECI). In a typical approach, ECIs are fitted to the first-principles calculated energy of a large number of ordered super-cells, through a variety of suggested procedures \cite{laks1992efficient, seko2009cluster, cockayne2010building, nelson2013compressive, seko2014cluster,leong2019robust, luis2021sparse, yang2022approaches, zhong2022ell_0}. Thermodynamic quantities can be obtained by sampling the CE energy with  Monte-Carlo simulations (CE-MC) \cite{binder1988monte,van2002self, van2002automating, van2002alloy}. This workflow allows fast statistical mechanics computation of configurational disorder, using only a relatively small number of first-principles calculations. More detailed descriptions of the CE-MC method can be found in various review papers \cite{van2013methods, wu2016cluster, sanchez2017foundations, kadkhodaei2021cluster, yang2022approaches, junzhong2022perspective}.

CE-MC can be performed in a canonical ensemble or in a grand-canonical ensemble. In a canonical ensemble, the configuration states are sampled with a fixed composition of each species. Using the Metropolis-Hastings algorithm \cite{metropolis1953equation, hastings1970monte}, a typical Metropolis step involves the swapping of the species occupying two randomly chosen sites (canonical swap). In a grand-canonical ensemble, the states are sampled under fixed chemical potentials allowing the relative amounts of each species to vary. A Metropolis step in the grand-canonical ensemble usually replaces the occupying species on one randomly chosen site with another species (single-species exchange). Grand canonical simulations are the preferred approach for studying phase transition in solids, as the simulation cell is always in a single-phase state, and phase transitions are relatively easy to observe. In contrast, multiple phases can coexist in canonical simulations, giving a disproportionate influence to the interfacial energy between phases.

Single-species exchanges can be applied without issue when the species are all charge-neutral atoms. However, in an ionic system in which all the species carry charge, net zero charge needs to be maintained, essentially coupling allowed species exchanges. For simulating ionic liquids, various methods have been proposed such as: inserting and removing only charge-neutral combinations of ions\cite{valleau1980primitive}; performing single insertion or deletion while controlling the statistical average of the net charge equal to be zero\cite{allen1987computer,frenkel1996understanding}; or using an expanded grand-canonical ensemble\cite{hatch2019improving}. However, charge balance in lattice-model CE-MC with arbitrary complexity has not been addressed in the literature yet.

In this study, we introduce two CE-MC sampling methods to handle the charge-balance constraint in the grand-canonical CE-MC for ionic systems with charge decoration. The first is the table-exchange (TE) method, in which MC samples are kept charge-neutral by using charge conserving multi-species exchanges. The second is the square-charge bias (SCB) method, which combines single-species exchanges with a penalty on the net charge to drive the system towards zero charge. We benchmark the computational efficiency over hyper-parameters in a complex rocksalt system with configurational disorder, and demonstrate proper usage strategies of these sampling methods. 

\begin{figure*}[ht]
    \centering
    \includegraphics[width=\linewidth]{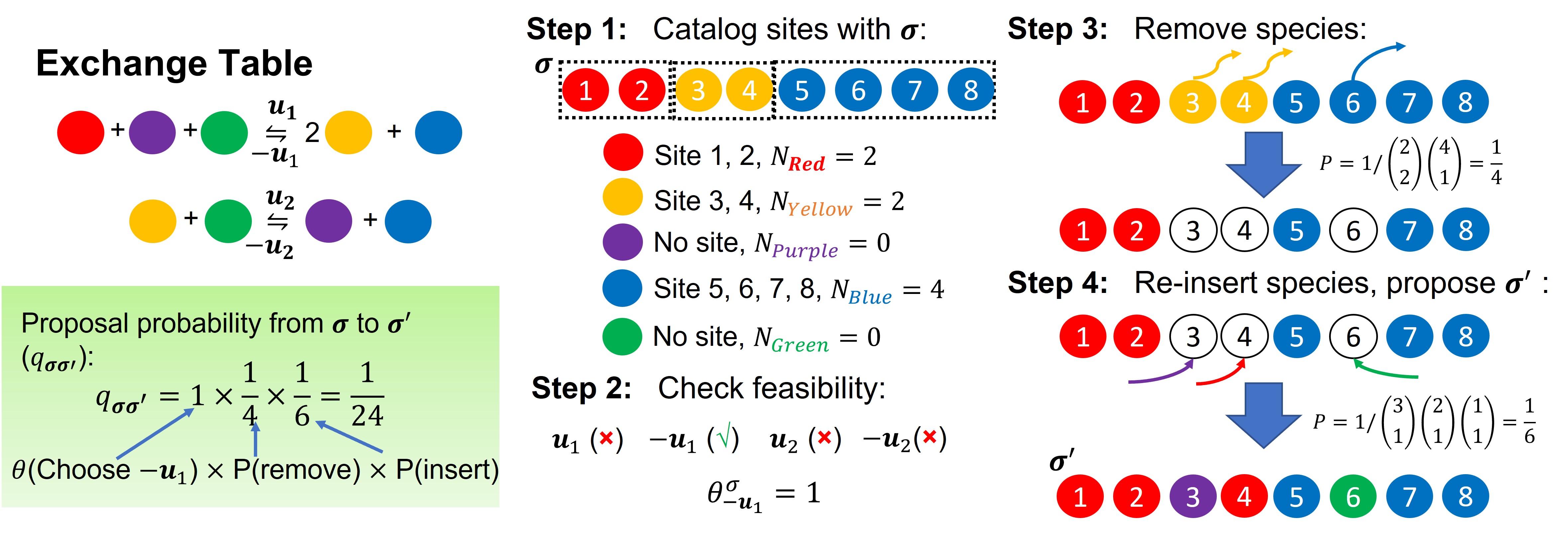}
    \caption{The procedure for proposing a table exchange in a conceptual quinary system. The system contains five different species (colored circles) and eight sites in a single sub-lattice (labeled with indices). Four exchange directions $\mathbf{u}_1$, $-\mathbf{u}_1$, $\mathbf{u}_2$ and $-\mathbf{u}_2$ are included in the exchange table. In the green box to the lower-left, the probability for proposing a particular configuration $\mathbf{\sigma}'$ from $\mathbf{\sigma}$ ($q_{\mathbf{\sigma\sigma}'}$) is calculated as the product of three probabilities: the probability for selecting an exchange direction, the probability for choosing removed species, and the probability for inserting new species to empty sites.}
    \label{fig:algo}%
\end{figure*}

\section{Methods}
For simplicity, the formalism in the following discussion is limited to materials with a single sub-lattice. However, the methodology can be easily extended to multiple sub-lattices. We also limit our investigation to the application of a charge-balance constraint; although more generic integral constraints on the composition (e.g., fixing the atomic ratio between particular components to follow a specific hyper-plane in the composition space) can be addressed in a similar manner.

\subsection{Table-exchange method}\label{sec:table-exchange}
In the grand-canonical ensemble with species carrying charge, every possible occupancy state must satisfy the following constraints:
\begin{equation}\label{eq:diop1}
    \begin{aligned}
    \sum_{s=1}^{S} C_s n_s &= 0,\\
    \sum_{s=1}^{S} n_s &= N,\\
    n_s \in \mathbb{N}&,\ \forall s \in \{1,2,\cdots,S\}
    \end{aligned}
\end{equation}
where $s$ is the label of a species, $n_s$ is the amount of species $s$ in configuration $\bm{\sigma}$, and $N$ is the total number of sites in the system. The first equation is a charge-balance constraint, where $C_s$ is the charge of species $s$. The second equation requires the number of species to be equal to the number of sites. Equation \ref{eq:diop1} is a system of linear Diophantine equations with natural number solutions. All integral solutions $\bm{n}=(n_1,\cdots,n_S)$ to these Diophantine equations can be represented as a bounded fraction of a $(S-2)$-dimensional integer grid in $\mathbb{N}^S$ \cite{robinson2001handbook}, specified as follows:

\begin{equation}
    \begin{aligned}
    \bm{n} &= \bm{n}_0 + \sum_{i=1}^{S-2} x_i \bm{v}_i,\\
    \text{s.t.}\ \ &x_i \in \mathbb{Z},\ \bm{v}_i\ \in \mathbb{Z}^S  \\
    &n_s \in \mathbb{N},\ n_s \leq N
    \end{aligned}
\label{eq:diop2}\end{equation}
where $\bm{n}_0$ is a base integer solution to Equation \ref{eq:diop1}, the $\bm{v}_i$'s are $S-2$ linearly independent basis vectors, and the $x_i$'s are integer coordinates on the grid.

Any vector $\bm{u}=\bm{n}'-\bm{n}$ pointing from one solution ($\bm{n}$) on the integer grid to another solution ($\bm{n}'$) is called an exchange direction. An exchange direction physically represents a composition transfer under the charge-balance constraint. A selected set $V$ among all possible $\bm{u}$ is called an exchange table. Based on the exchange table $V$, we can define a random walk process between charge-balanced compositions as follows:

\textbf{(1)} Using the current composition $\bm{n}$, select one direction $\bm{u}$ from all feasible directions in the predefined exchange table $V$. The feasibility of a direction $\bm{u}$ is defined with the requirement, that for all $u_s < 0$ (i.e. species $s$ is being removed), we have $n_s > -u_s$, ensuring a move towards direction $\bm{u}$ would not result in a negative amount of any species.

\textbf{(2)} Perform the operation to the occupancy configuration according to the selected exchange $\bm{u}$, such that the composition $\bm{n}$ changes to $\bm{n}+\bm{u}$. Given $\bm{u}=(u_1, u_2, \cdots, u_S)$, one such operation can be achieved by removing $-u_s$ of species $s$ from the occupancy for all $u_s<0$; then inserting $u_s$ of species $s$ into the empty sites, for all $u_s > 0$. Such an operation is called a table exchange. It results in a simultaneous exchange of species on multiple sites and is always charge conserving. The number of sites $U$ to be exchanged is called the exchange size in direction $\bm{u}$. Because any exchange should conserve the site number, $\sum_{s} u_s=0$. Therefore,  $U=\sum_{u_s>0}{u_s}=\sum_{u_s<0}{-u_s}$.

A complete exchange table should have ergodicity, which means an MC simulation should be able to reach any charge-balanced composition from an arbitrary starting configuration. Once ergodicity is satisfied, the number of sites involved in the exchange directions should be minimal, as exchanging a large number of sites in a Metropolis step can lead to low acceptance ratio and thus inefficient sampling of the configuration space. It is not necessary, nor practical, to include all possible directions $\bm{u}$ in the table. Usually, as a minimal setup, one can choose $S-2$ linearly independent basis vectors ($\{\bm{v}_i\}$) with minimal exchange size as well as their inverse vectors ($\{-\bm{v}_i\}$). The ergodicity of a table can be checked by enumerating charge balanced compositions in a specific super-cell size as vertices of a graph, and checking graph connectivity between the compositions using vectors in the table as the edges of the graph. If ergodicity is not satisfied with the minimal setup, and the unreachable compositions are of interest, vectors linking the disconnected composition to other compositions should be added to the table, until the ergodicity is guaranteed.

According to the statements above, given an exchange table $V$, one can propose grand-canonical Metropolis steps using the following procedure, as illustrated schematically in Figure \ref{fig:algo}:

\begin{enumerate}
    \item Create a catalog of sites in the lattice. For a starting occupancy state $\bm{\sigma}$, indices $j$ of sites are grouped by their occupied species $s$, to create sets $J_s = \{j|\sigma_j=s\}$.
    \item Select one feasible direction $\bm{u}$ from table $V$. The subset of table $V$ with all feasible directions at occupancy $\bm{\sigma}$ is denoted as $V_{\bm{\sigma}}$, and the probability for selecting direction $\bm{u}$ is denoted as $\theta_{\bm{u}}^{\bm{\sigma}}$. In this work, we select all feasible directions at an equal probability ($\theta_{\bm{u}}^{\bm{\sigma}}=1/\mathrm{card}(V_{\bm{\sigma}}), \forall \bm{u} \in V_{\bm{\sigma}}$).
    \item For all $u_s<0$, randomly pick $-u_s$ sites from catalog $J_s$ without replacement. Select all possible picking combinations at equal probability ($P=1/\prod_{u_s < 0}\binom{n_s}{-u_s}$). Remove the species from selected sites.
    \item For all $u_s>0$, randomly select $u_s$ empty sites from the $U$ empty sites created in Step 3 without replacement, and insert species $s$ back to selected empty sites. All possible combination of choices can be selected with equal probability ($P=\prod_{u_s>0}u_s!/U!$). Propose the resulting occupancy state $\bm{\sigma}'$.
\end{enumerate}

Note that the procedure above can result in an asymmetry between the exchange proposal probability from $\bm{\sigma}$ to $\bm{\sigma}'$ and the inverse proposal probability from $\bm{\sigma}'$ back to $\bm{\sigma}$. Such a proposal asymmetry can be balanced by multiplying with a composition dependent importance factor to adjust the acceptance probability as given by Equation \ref{eq:importance}, such that detailed balance is ensured and the correct distribution is reached (see Supplementary Information for a detailed derivation).

\begin{widetext}
\begin{equation}
    p_{\bm{\sigma \sigma'}} = \min \left\{1,\  \frac{\theta^{\bm{\sigma'}}_{-\bm{u}} \prod_{u_s \neq 0} n_s !}{\theta^{\bm{\sigma}}_{\bm{u}} \prod_{u_s \neq 0} (n_s + u_s)!}
    \exp \left[-\frac{1}{k_B T} \left(\Delta E_{\bm{\sigma \sigma'}} - \sum_{s} \mu_s u_s\right) \right]
    \right\}
\label{eq:importance}
\end{equation}
\end{widetext}

In addition to table exchanges which change the composition, a portion $(0 \leq w < 1)$ of canonical swaps can also be mixed in the proposal. These canonical swaps can directly transfer between occupancies under the same composition with much less computational cost than table exchanges and are added to help explore occupancies with the same composition more efficiently, rather than having to do so with a combination of table exchanges. In the discussion section, we will illustrate the importance of hyper-parameter $w$ in the TE method.

\subsection{Square-charge bias method}
Compared with single-species exchanges, proposing a table-exchange step and computing its energy change is more time-consuming. It is still desirable to find a method using single-species exchanges that still conserves charge-balance. In the square-charge bias (SCB) method, we use single-species exchanges to span all occupancies regardless of charge-balance. States in the Markov chain are allowed to leave charge-balance. However, we add a penalty on the square of the net charge to the Hamiltonian to drive the sampled configurations back to charge-balance. The acceptance probability of each single-species exchange step is evaluated using the following penalized Hamiltonian:
\begin{equation}
H_{\mu,\lambda}(\bm{\sigma}) = E(\bm{\sigma}) - \sum_{s} \mu_s n_s + \lambda k_B T C(\bm{\sigma})^2
\label{eq:penalty}\end{equation}
where $E(\bm{\sigma})$ is the energy of occupancy $\bm{\sigma}$ computed from CE. The charge penalty factor $\lambda > 0$ is a hyper-parameter in the SCB method, and $k_B T$ is included in the penalty to make $\lambda$ dimensionless. $C(\bm{\sigma})$ is the net charge of occupancy $\bm{\sigma}$:
\begin{equation}
C(\bm{\sigma}) = \sum_s C_s n_s
\end{equation}

In a SCB run, we start from a charge-balanced state. After reaching thermal equilibration, from all states in the equilibrated sample, we compute the average of physical quantities with only charge-balanced states (i.e., states with $C(\sigma)=0$). The charge-balance constraint is rigorously satisfied in our estimation of the sample in this manner. Meanwhile, when $C(\bm{\sigma})=0$, we always have $H_{\mu,\lambda}(\sigma) = E(\bm{\sigma}) - \sum_s \mu_s n_s$. Therefore, the true grand-canonical distribution should also be recovered among the charge-balanced sample states. The effect of hyper-parameter $\lambda$ on SCB is demonstrated in the discussion section

\subsection{Comparing computational efficiency of sampling methods}
If a CE-MC algorithm has hyper-parameters, it is desirable to optimize them such that the thermodynamic properties can be estimated accurately, with minimal computational cost. To estimate the ensemble average ($\overline{\theta}$) of a physical quantity $\theta$, we run CE-MC and generate a Markov chain of states, and at each step $p$ the value of $\theta$ for the current configuration is recorded as $\theta_p$. We denote $\overline{\theta}_{[p,q]}$ as the mean of $\theta$ in a block from step $p$ to step $q$. In the SCB case, block means are taken from only charge-balanced states in the block. After thermal equilibration, we define the variance of $\overline{\theta}_{[p, p+L]}, \overline{\theta}_{[p+L, p+2L]}, ...$ as the block mean variance ($\mathrm{Var}(\overline{\theta}_L)$) at block length $L$. The block mean variance can be used as a measure of uncertainty, if we estimate $\overline{\theta}$ with one of the block means above.

Suppose the true variance of $\theta$ is $\tau^2$ in the ensemble, then the sampling efficiency on property $\theta$ can be defined as follows\cite{gelman1996efficient}:
\begin{equation}\label{eq:neff}
    \mathrm{eff}(\theta) = \frac{\tau^2}{L\mathrm{Var}(\overline{\theta}_L)}
\end{equation}
With ideal independent random sampling, one can expect $\mathrm{Var}(\overline{\theta}_L) = \tau^2/L$, such that $\mathrm{eff}(\theta)=1$. In reality, Metropolis steps are always correlated and the efficiency is expected to be lower than 1 ($\mathrm{eff}(\theta)<1$). A CE-MC algorithm with higher sampling efficiency is less correlated and can thus reduce the uncertainty of estimation to a low level with fewer sampling steps. In the TE method, the time cost of a table exchange is significantly higher than a canonical swap, such that counting the number of Metropolis steps no longer accurately reflects the computational cost. In this work, we used a modified version to evaluate the sampling efficiency. We replace the block length $L$ in Equation \ref{eq:neff} with $\overline{T}_L$, which is the average CPU time spent in each block.
\begin{equation}
    \mathrm{eff}_t(\theta) = \frac{1}{\overline{T}_L\mathrm{Var}(\overline{\theta}_L)}
    \label{eq:teff}
\end{equation}
We define $\mathrm{eff}_t(\theta)$ in Equation \ref{eq:teff} as the computational efficiency on the property $\theta$, and use it for benchmarking the algorithm under varied hyper-parameters.

\section{Numerical Results}
\begin{figure*}[htb]
    \centering
    \includegraphics[width=0.9\linewidth]{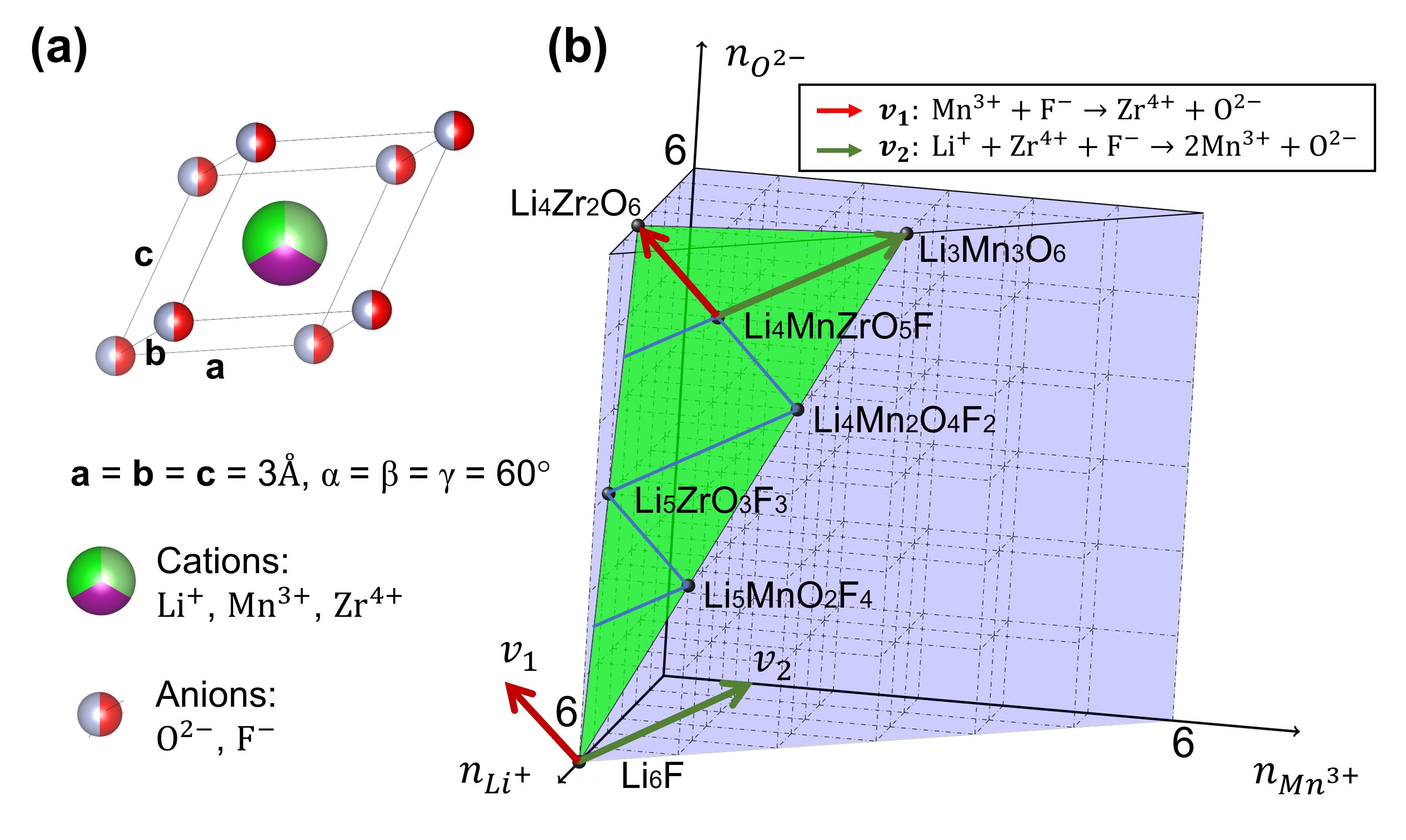}
    \caption{Primitive cell and exchange directions in the LMZOF disordered rocksalt. (a) Rocksalt primitive cell of LMZOF, with partial occupancies of $\mathrm{Li}^{+}$, $\mathrm{Mn}^{3+}$, $\mathrm{Zr}^{4+}$ on the cation sub-lattice and $\mathrm{O}^{2-}$, $\mathrm{F}^{-}$ on the anion sub-lattice. (b) Compositions of LMZOF in a super-cell of size 6 (6 cation sites and 6 anion sites). The x-, y-, and z- axis represent the amount of $\mathrm{Li}^{+}$, $\mathrm{Mn}^{3+}$ and $\mathrm{O}^{2-}$, respectively. The amount of $\mathrm{Zr}^{4+}$ and $\mathrm{F}^{-}$ can be computed by satisfying site number conservation on the cation and the anion sub-lattices. The purple dashed grid in three dimensions includes arbitrary compositions without enforcing charge-balance. The solid grid on the green plane includes charge-balanced compositions only. Basis vectors $\bm{v}_1$ and $\bm{v}_2$ are marked with dark green and red arrows, respectively. The reaction formulas corresponding to $\bm{v}_1$ and $\bm{v}_2$ are listed on the top right. The inverse directions are not displayed.}
    \label{fig:diop}%
\end{figure*}

In this section, we demonstrate the influence of the hyper-parameters on the computational efficiency and thermal equilibration in the TE and SCB methods. 
We performed CE-MC simulations in a disordered rocksalt system. A rocksalt crystal structure is a basic prototype of ionic materials consisting of an FCC cation and an FCC anion sub-lattice, mimicking the basic chemistry of some novel Li-ion cathode systems which have been modeled with CE-MC in recent studies \cite{lun2021cation, McColl2022oxy_redox,guo2022intercalation}. In our system, $\mathrm{Li}^{+}$, $\mathrm{Mn}^{3+}$, $\mathrm{Zr}^{4+}$ are distributed on the cation sub-lattice, and $\mathrm{O}^{2-}$, $\mathrm{F}^{-}$ are present on the anion sub-lattice. We refer to this system as LMZOF, with the primitive cell presented in Figure \ref{fig:diop}(a) and the exchange directions shown in Figure \ref{fig:diop}(b).

For the TE and SCB methods, we performed simulations under various hyper-parameters $w$ and $\lambda$. After thermal equilibration, we calculated the computational efficiencies (Equation \eqref{eq:teff}) for the following quantities: (1) $E$ (configurational energy per super-cell), (2) $x_{\mathrm{LiMnO}_2}$ (atomic percentage of $\mathrm{LiMnO}_2$) and (3) $x_{\mathrm{Li}_2\mathrm{ZrO}_3}$ (atomic percentage of $\mathrm{Li}_2\mathrm{ZrO}_3)$. To discuss how the hyper-parameters $w$ and $\lambda$ affect the computational efficiency and thermal equilibration in the TE and SCB methods, we designed two simulation experiments: (1) T= 5000 K to simulate the system in a state of complete solubility and (2) T = 2000 K to simulate the system in a single phase ($\mathrm{Li}_2\mathrm{ZrO}_3$). In experiment (1), the sampling efficiencies were plotted as a function of hyper-parameters $w$ and $\lambda$. In experiment (2), the thermal equilibration process was demonstrated with simulation trace plots, which showed the value of thermodynamic properties (such as the composition and the configuration energy) as a function of simulation step. The details of these simulations are provided in the Supplementary Information.

\subsection{Simulation with table exchange}
\begin{figure*}[htb]
    \centering
    \includegraphics[width=\linewidth]{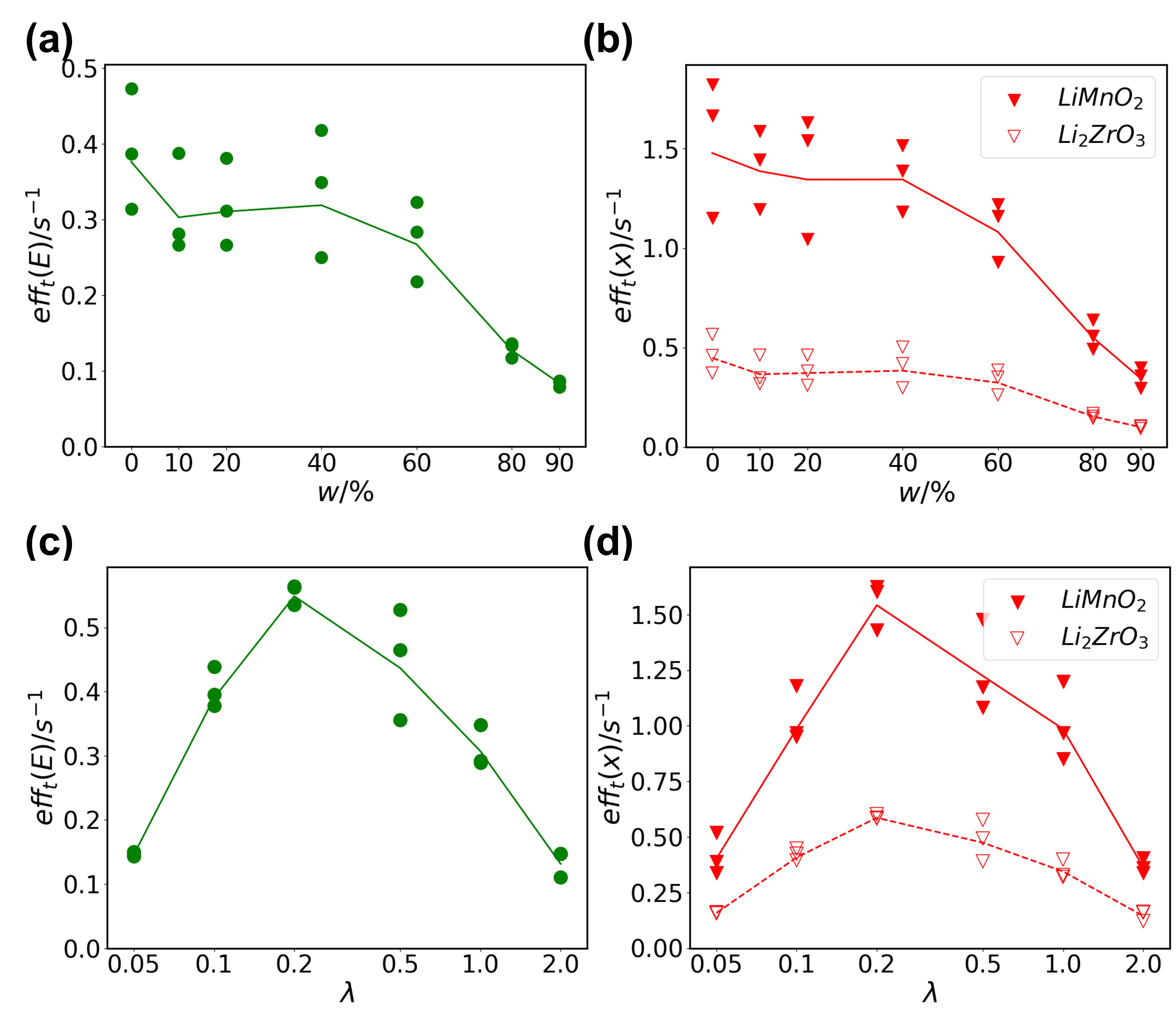}
    \caption{TE and SCB computational efficiencies in the LMZOF system at T =5000 K as a function of the canonical swap percentage $w$ and the charge penalty factor $\lambda$. For each $w$ and each $\lambda$, three simulations were run starting from different initial states. The average of three measurements for each $w$ and each $\lambda$ are connected with lines. (a) TE computational efficiencies for energy ($\mathrm{eff}_t(E)$, green dots and line) as a function of $w$. (b) TE computational efficiencies for sampling the $\mathrm{LiMnO}_2$ composition ($\mathrm{eff}_t(x_{\mathrm{LiMnO}_2})$, red solid triangles and solid line) and the  $\mathrm{Li}_2\mathrm{ZrO}_3$ composition ($\mathrm{eff}_t(x_{\mathrm{Li}_2\mathrm{ZrO}_3})$, red hollow triangles and dashed line) as a function of $w$. (c) SCB computational efficiency for energy ($\mathrm{eff}_t(E)$, green dots and line) as a function of $\lambda$. (d) SCB computational efficiency for sampling the $\mathrm{LiMnO}_2$ composition ($\mathrm{eff}_t(x_{\mathrm{LiMnO}_2})$, red solid triangles and solid line) and the $\mathrm{Li}_2\mathrm{ZrO}_3$ composition ($\mathrm{eff}_t(x_{\mathrm{Li}_2\mathrm{ZrO}_3})$, red hollow triangles and dashed line) as a function of $\lambda$.}
    \label{fig:effs}%
\end{figure*}

In the TE method, the parameter $w$ tunes the ratio of table exchanges to canonical swaps, where only table exchanges can explore different compositions. Most physical systems have a critical temperature $T_c$(or a series of critical temperatures) below which they phase separate into phases of distinct compositions (compounds or elemental states). Above $T_c$ complete solubility can be found. Under such circumstances, a low $w$ will include more table exchanges to explore a wide distribution of compositions, and gives better sampling efficiencies. Figure \ref{fig:effs} (a) and (b) shows the TE computational efficiency under 5000 K, where all the components in LMZOF are fully miscible (see Supplementary Information). The computational efficiencies for the configurational energy and compositions are both maximized at $w=0\%$, indicating that no canonical swaps should be included.

\begin{figure*}
    \centering
    \includegraphics[width=\linewidth]{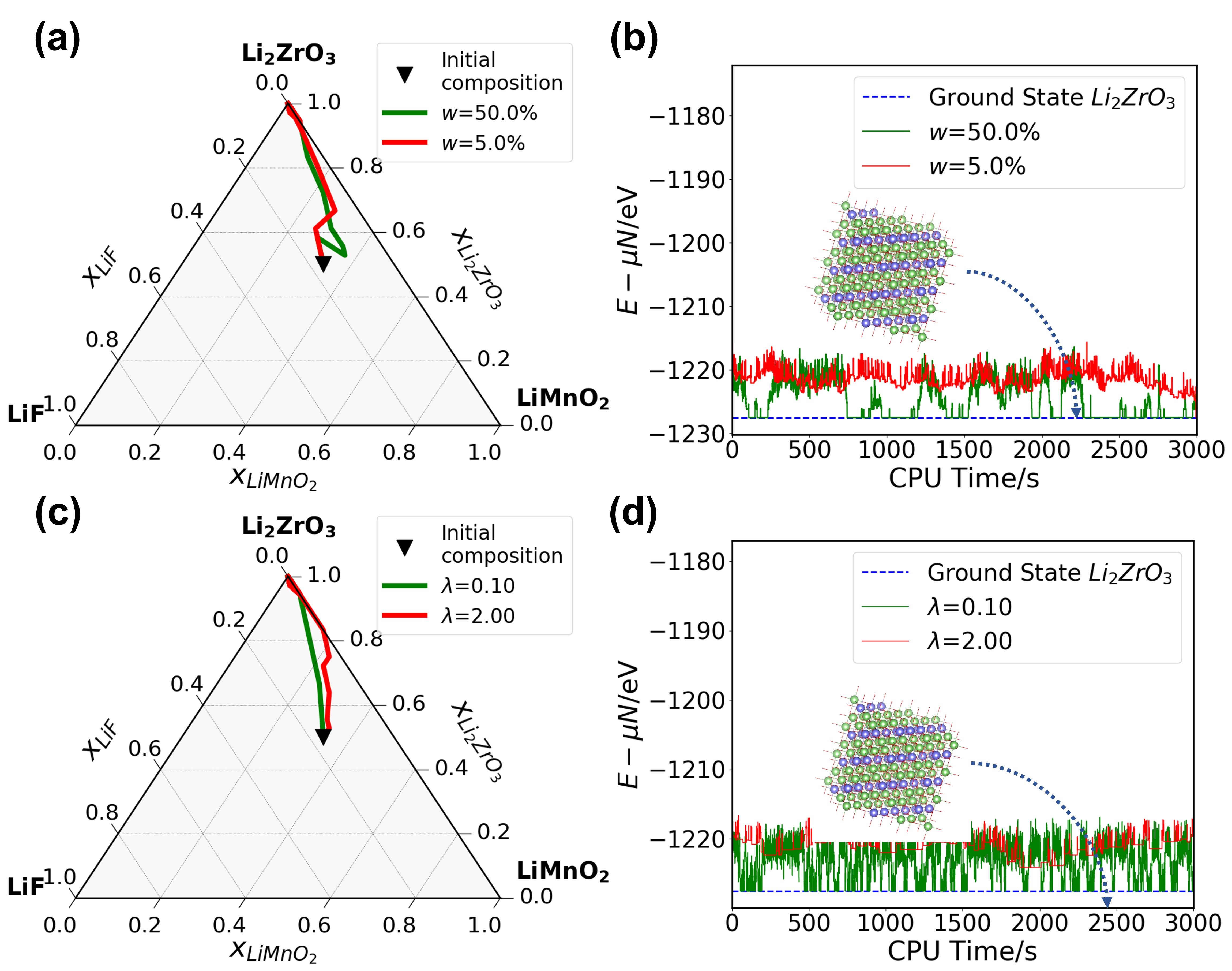}
    \caption{Trace plots of TE simulations in LMZOF system, at T = 2000 K, $w=5\%$ and $50\%$; and SCB simulations at T = 2000 K, $\lambda=0.1$ and $\lambda=2.0$. The simulations started from the same occupancy configuration. In (a) and (c), the simulated trajectory of compositions are plotted in the LMZOF phase space, and the initial state composition is marked with a black triangle. In (c) and (d), the simulated trajectories of the energy ($E-\mu N$) are plotted as a function of the simulation time. The blue dashed baseline shows the energy of the $\mathrm{Li}_2\mathrm{ZrO}_3$ ground state.  (a) Simulated trajectory of composition using $w=5\%$(red) and $50\%$(green) in TE.  (b) Simulated trajectory of energy with the chemical potential subtracted ($E-\mu N$), using $w=5\%$(red) and $w=50\%$(green) in TE. (c) Simulated trajectory of compositions using $\lambda=2.0$(red) and $0.1$(green) in SCB. (d) Simulated trajectory of energy with the chemical potential subtracted ($E-\mu N$), using $\lambda=2.0$(red) and $\lambda=0.1$(green) in SCB.}
    \label{fig:traces}%
\end{figure*}

Nevertheless, it is not always safe to fully exclude canonical swaps. Below the critical temperature, the grand-canonical ensemble distribution is usually concentrated near the composition of a single phase; thus, the ability to explore different occupancies with the same composition is more important (namely, the ability of canonical state transfers). With only table exchanges, it is still possible to achieve a canonical transfer by performing multiple exchanges when the sum of all exchange directions equals to zero. However, besides being computationally more expensive than a canonical swap, table exchanges perturb many sites simultaneously and are therefore more likely to propose energetically unfavorable configurations, resulting in lower acceptance ratio. As a result, having too low of a canonical swap percentage $w$ can reduce the computational efficiencies, and lead to slow thermal equilibration, especially at a relatively low temperature. Such an example is illustrated in Figure \ref{fig:traces} (a) and (b) in LMZOF at 2000 K. Even though the simulation was able to reach equilibrium at a single phase composition ($\mathrm{Li}_2\mathrm{ZrO}_3$, Figure \ref{fig:traces}(a)) for $w=5\%$ (red), compared to $w=50\%$ (green), it failed to equilibrate to the correct ground-state configuration (the layered structure, shown in Figure \ref{fig:traces}(b)) within a time limit of 3000 seconds. 

\subsection{Simulation with square-charge bias}
In the SCB method, the penalty factor $\lambda$ controls the trade-off between the fraction of charge-balanced states in the Markov chain and the Metropolis acceptance probability. Figure \ref{fig:effs} (c) and (d) show the sampling efficiency in LMZOF at 5000 K. Optimal efficiency is found at an intermediate $\lambda$ value ($\lambda=0.2$). When the penalty $\lambda$ is too small, the simulation can wander too far from charge-balance, and barely revisits charge-balanced configurations. A too large $\lambda$ value limits low-barrier pathways towards new charge balanced configurations. Near either of these extreme circumstances, the sampling efficiency of SCB decreases. Figure \ref{fig:traces} (c) and (d) show at T = 2000 K an overly large charge penalty $\lambda=2.0$ (red) causes slow configurational equilibration to the layered $\mathrm{Li}_2\mathrm{ZrO}_3$ ground state because of the aforementioned limitation to low-barrier pathways.

In the SCB approach, we can define the occupancy transfer rate ($r_o$) and the composition transfer rate ($r_c$) as follows:
\begin{equation}
\begin{aligned}
    r_o &= \frac{\mathrm{Count\ of\ occupancy\ transfers}}{\mathrm{CPU\ time\ elapsed}}\\
    r_c &= \frac{\mathrm{Count\ of\ composition\ transfers}}{\mathrm{CPU\ time\ elapsed}}
\end{aligned}
\end{equation}
where an occupancy transfer is counted when the Markov chain arrives at a new charge-balanced occupancy different from the last recorded charge-balanced state, while a composition transfer is counted when a charge balanced composition different from the last record is reached.

\begin{figure}
    \centering
    \includegraphics[width=\linewidth]{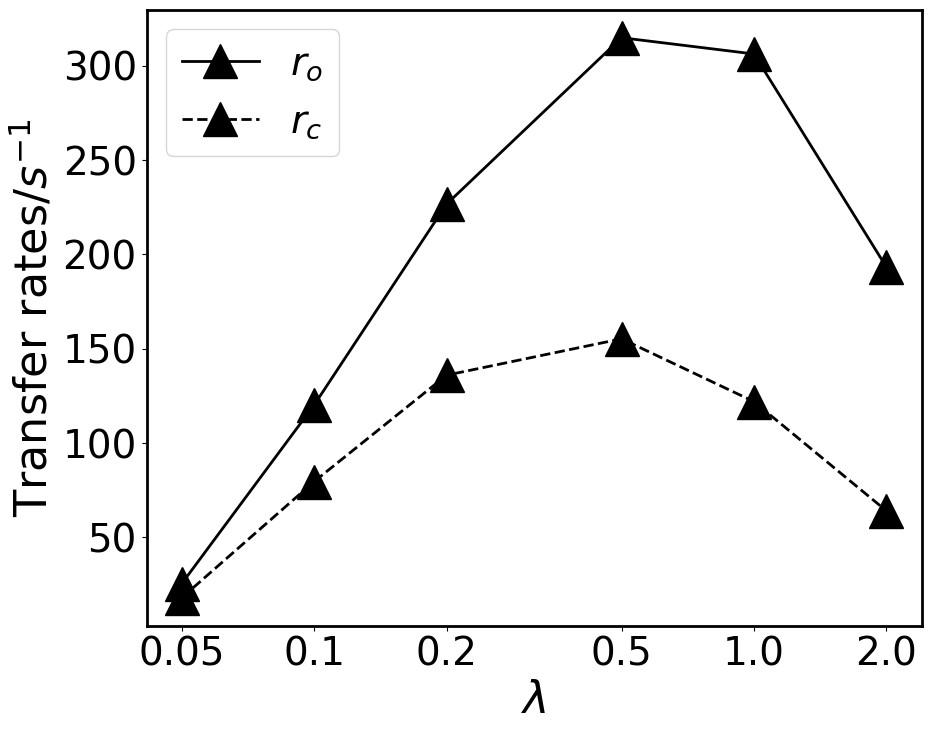}
    \caption{Occupancy ($r_o$, solid line) and composition ($r_c$, dashed line) transfer rates in SCB simulations of LMZOF under 5000 K, using varied $\lambda$.}
    \label{fig:5000k_rates}%
\end{figure}

In Figure \ref{fig:5000k_rates}, we computed the transfer rates in the SCB simulations at 5000 K in LMZOF. The maximum transfer rates occur at $\lambda=0.5$. When compared to the sampling efficiency trend in Figure \ref{fig:effs} (c) and (d), the efficiencies at $\lambda=0.5$ are only $20\%$ lower than the optimal sampling efficiency taken at $\lambda=0.2$. Compared to the computational efficiency, the transfer rates can be tracked step by step without waiting for multiple blocks of the Markov chain to complete. They can also give a satisfactory estimation to the optimal $\lambda$. Therefore, when using the SCB method, one may instead choose an optimal $\lambda$ to maximize the transfer rates as a alternative to maximizing the computational efficiency.

\section{Discussion \& Summary}
We introduced two methods to perform grand-canonical CE-MC simulation with a charge-balance constraint, enabling thermodynamic calculations for ionic materials with configurational disorder.

The effect of the fraction of canonical exchanges $w$ mixed into the grand canonical trajectory, and the charge penalty factor $\lambda$ in the TE and SCB methods are presented. In the TE method, using a proper $w$ is essential to efficiently explore and equilibrate among same-composition configurations. In the SCB method, the penalty factor $\lambda$ controls the trade-off between the ability to revisit charge balance and the ability to leave charge-balance to explore new states. We show that the hyper-parameters $w$ and $\lambda$ can be optimized to achieve a satisfactory computational efficiency.

In addition, as illustrated in Figure \ref{fig:effs}, the maximum computational efficiencies of the TE method and the SCB method are close ($\mathrm{eff}_t(E) \approx 0.5 s^{-1}$, $\mathrm{eff}_t(x_{\mathrm{LiMnO}_2}) \approx 1.5 s^{-1}$ and $\mathrm{eff}_t(x_{\mathrm{Li}_2\mathrm{ZrO}_3} \approx 0.5 s^{-1}$). Therefore, in a system with small table-exchange sizes (e.g., in LMZOF, $U \leq 3$), the TE method is shown to have similar performance as the SCB approach. However,when the exchange table includes large-sized exchanges, the sampling efficiency of the TE method can be limited. Consider a disordered rocksalt-like system in chemical space of $x\cdot\mathrm{LiF} +(1-x)\cdot\mathrm{Li}\mathrm{Ni}^{2+}_{1/3}\mathrm{Mn}^{3+}_{1/3}\mathrm{Ti}^{4+}_{1/3}\mathrm{O}_2$ ($0 \leq x \leq 1$, referred as LNMTOF). The system consists of $\mathrm{Li}^{+}$, $\mathrm{Ni}^{2+}$, $\mathrm{Mn}^{3+}$, $\mathrm{Ti}^{4+}$ on the cation sub-lattice, and $\mathrm{O}^{2-}$, $\mathrm{F}^{-}$ on the anion sub-lattice, with an additional requirement that $n_{\mathrm{Ni}^{2+}}=n_{\mathrm{Mn}^{3+}}=n_{\mathrm{Ti}^{4+}}$. The minimal basis exchange table in LNMTOF contains the following exchanges ($U = 9$):
\begin{equation}
    3 \mathrm{Li}^{+} + 6 \mathrm{F}^{-} \Longleftrightarrow 
    \mathrm{Ni}^{2+} + \mathrm{Mn}^{3+} + \mathrm{Ti}^{4+} + 6 \mathrm{O}^{2-}.
\end{equation}

\begin{figure}
    \centering
    \includegraphics[width=\linewidth]{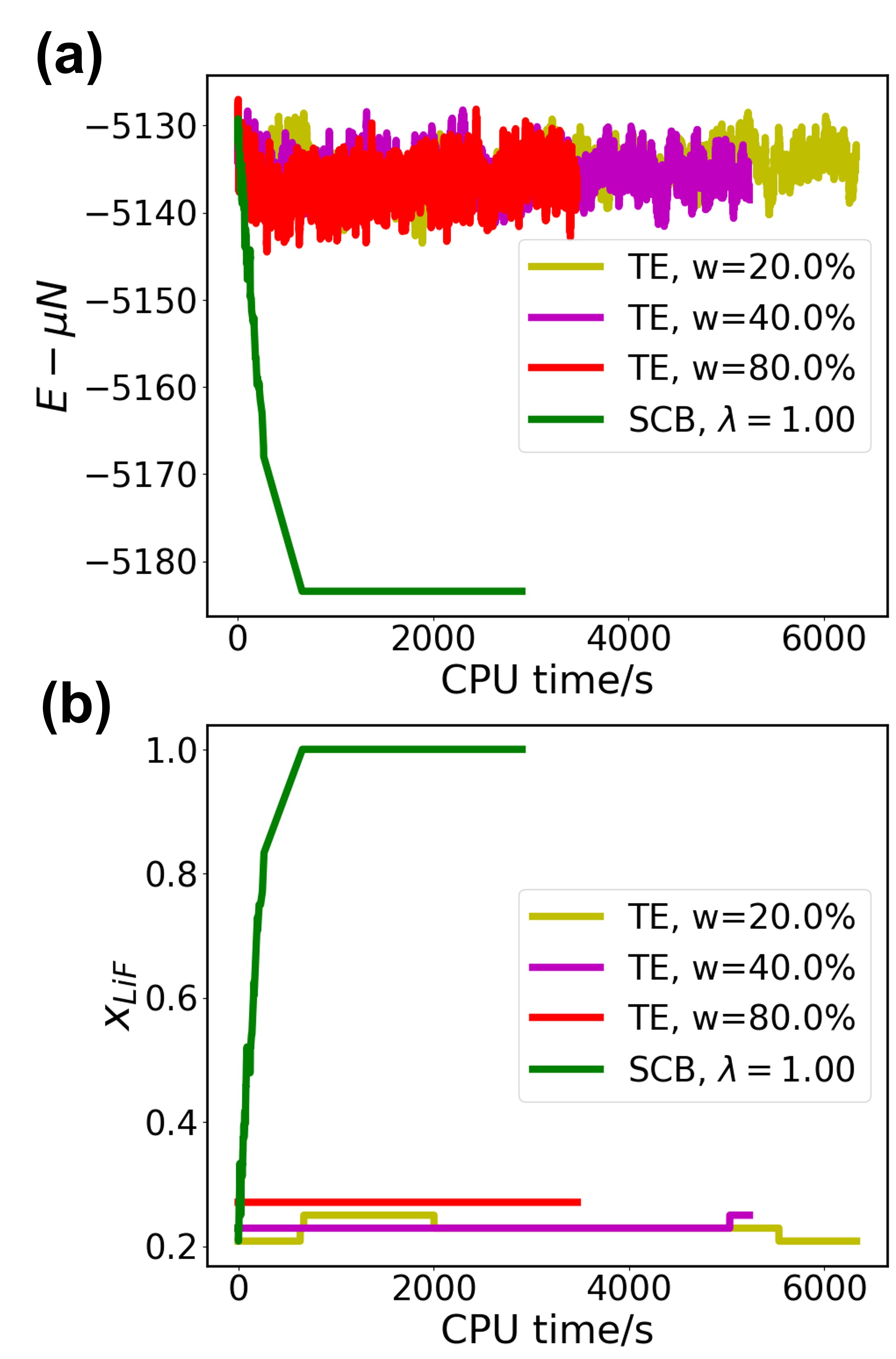}
    \caption{Trace plots of TE simulations in LNMTOF system at T = 1600 K, $w=20\%$ (yellow), $40\%$ (purple) and $80\%$ (red); and of an SCB simulation at the same temperature and $\lambda=1.0$ (green). (a) Simulated trajectory of the energy ($E-\mu N$), as a function of the simulation time. (b) Simulated trajectory of the $\mathrm{LiF}$ composition ($x_{\mathrm{LiF}}$) as a function of the simulation time.
    }
    \label{fig:lmntof_eq}%
\end{figure}

\begin{figure}
    \centering
    \includegraphics[width=\linewidth]{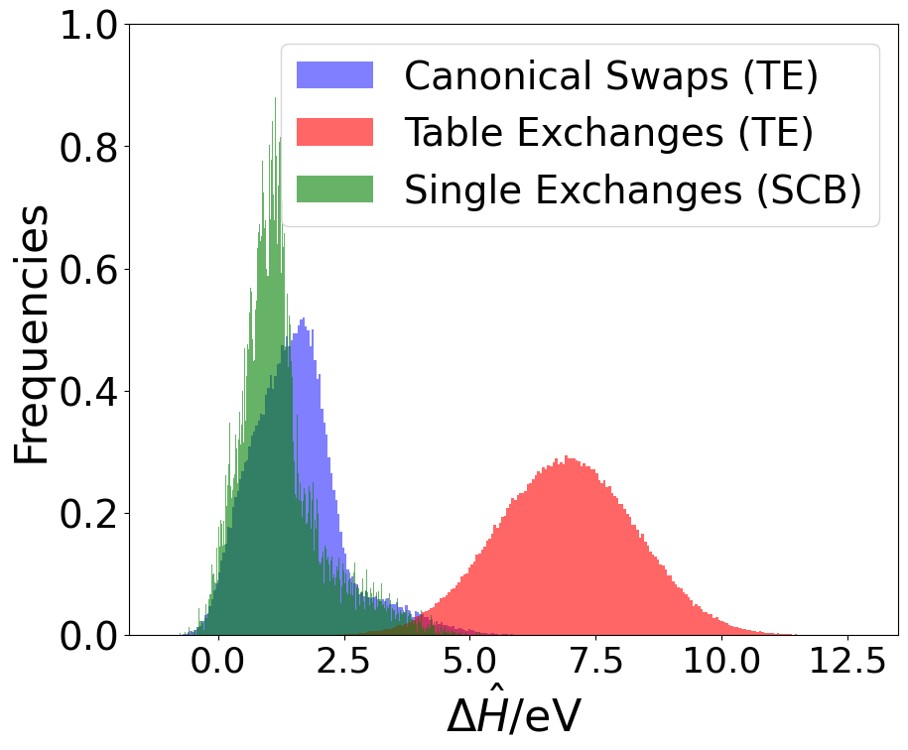}
    \caption{Distribution of the effective perturbation energy ($\Delta \hat{H}$) of three types of Metropolis steps: canonical swaps (blue), table exchanges (red) and single exchanges (green). Metropolis steps were applied to 10 snapshot LNMTOF configurations drawn from the TE simulation at T = 1600 K, $w=40\%$. In canonical swaps, $\Delta \hat{H}=\Delta E$. In table exchanges, $\Delta \hat{H}=\Delta (E - \mu N)$. In single-species exchanges (SCB), $\Delta \hat{H}=\Delta (E - \mu N + \lambda k_B T C^2)$, where T = 1600K and $\lambda=1.0$.
    }
    \label{fig:lmntof_barriers}%
\end{figure}

Figure \ref{fig:lmntof_eq} (a) and (b) shows the trajectories of energy subtracted by chemical potentials ($E-\mu N$) and $\mathrm{LiF}$ atomic percentage ($x_{LiF}$) simulated at T=1600K, using the TE method with varied $w$ and the SCB method with $\lambda=1.0$ (see details in Supplementary Information). Regardless of the value of $w$, all TE simulations are unable to reach the ground-state $LiF$ as suggested by the SCB simulation as the transfers between compositions are nearly prohibited, suggesting very low acceptance ratio of table exchanges. Ten random configurations were drawn as snapshots from the Markov chain generated by the TE simulation at $w=40\%$, to which three types of Metropolis steps (canonical swaps, table exchanges, single exchanges) were applied to calculate the effective perturbation energies ($\hat{H}$). The distributions of $\hat{H}$ with each type of Metropolis steps are shown in Figure \ref{fig:lmntof_barriers}. The table exchanges (red) in LNMTOF show significantly higher perturbation energy compared to the canonical swaps (blue) and single exchanges in the SCB method (green). This is because many sites are required to exchange simultaneously. The large energy perturbation in TE prohibits effective transfer between different compositions, and explains the slow thermal equilibration in the TE method. We suggest using the SCB method instead of the TE method for acceptable efficiency of thermal equilibration when large-sized exchanges are included.

% Change to 1.0\linewidth in reprint.

In summary, we recommend the following strategy to apply TE and SCB in practical CE-MC calculations:
\begin{enumerate}
    \item Choose the proper method according to the size of table exchanges (based on the exchange size $U$). When the size of table exchange is large (for example, $U > 4$), TE should be used cautiously as it may lead to low sampling efficiency and slow thermal equilibration.
    \item Scan a series of $w$ or $\lambda$ coarsely to  benchmark the computational efficiency.  For example, a series of $w=90\%, 70\%, 50\%, 30\%$, and $10\%$; or a series of $\lambda = 0.1, 0.2, 0.5, 1.0$ and $2.0$ may be sufficient.
    \item Perform short trial simulations with each $w$ or $\lambda$ at the temperature and chemical potentials of interest. Record the trace of properties along with the CPU time elapsed. By inspecting the convergence of $E-\mu n$ and compositions, hyper-parameter values that results in slow thermal equilibration can be ruled out. Search among the remaining values of $w$ or $\lambda$, in order to maximize the computational efficiency ($\mathrm{eff}_t$).
    \item Continue the simulation with the optimal hyper-parameter value, and generate the formal MC samples.
\end{enumerate}

\section{Acknowledgement}
This work was funded by the U.S. Department of Energy, Office of Science, Office of Basic Energy Sciences, Materials Sciences and Engineering Division under Contract No. DE-AC02-05-CH11231 (Materials Project program KC23MP). The work was also supported by the computational resources provided by the Extreme Science and Engineering Discovery Environment (XSEDE), supported by National Science Foundation grant number ACI1053575; the National Energy Research Scientific Computing Center (NERSC), and the Lawrencium computational cluster resource provided by the IT Division at the Lawrence Berkeley National Laboratory. 

\bibliography{cn.bib}

\end{document}